\documentclass{kapproc} 
\let\footnote\savefootnote
\let\footnotetext\savefootnotetext 
\let\footnoterule\savefootnoterule 
\kluwerbib

\RequirePackage{graphicx}%
\RequirePackage{epsf}%
\RequirePackage{psfig}%

\begin{document}

\articletitle{X--Ray IGM in the Local Group}

\author{Andrew Rasmussen, Steven M. Kahn and Frits Paerels}

\affil{Columbia University}
\email{arasmus@astro.columbia.edu}


\begin{abstract}
Recent observations with the dispersive X-ray spectrometers aboard
{\it Chandra} and {\it Newton Observatory} have begun to probe the
properties of the X-ray intergalactic medium (IGM) at small
redshifts. Using large quantities ($\sim$950~ksec) of spectroscopic
data acquired using the Reflection Grating Spectrometer (RGS) aboard
{\it Newton Observatory}, we investigated the intervening material
toward three  low redshift, high Galactic latitude Active Galactic
Nuclei (AGNs) with nominally featureless spectra: Mrk~421, PKS~2155$-$304
and 3C~273. 
Each spectrum provides clear evidence for what appears to be a local
($z \sim 0$), highly ionized absorbing medium  betrayed by the
\ion{O}{7} 1s--2p resonance transition feature seen at 21.6\AA\
($\rm N_{\rm OVII} \sim 10^{16}\,\rm cm^{-2}$). Measurements are also
made for the Ly~$\alpha$ transition of the adjacent ionization state,
(\ion{O}{8}; 18.97\AA), which potentially constrains the absorber's 
temperature. Finally, in a collisional
equilibrium approximation, upper limits to diffuse emission
intensities place upper limits on the electron density ($n_{\rm e} <
2\times 10^{-4}\,\rm cm^{-3}$), lower limits on the scale length of
the absorber ($\rm L > 140\,\rm kpc$) and lower limits on its mass
($\rm M > 5\times 10^{10}\,\rm M_\odot$). Limits on the absorber's
scale length and its velocity distribution lead us to identify it with
the Local Group. Having detected the hot gas in our Local Group in
absorption, it should be feasible to detect also the extended
structure of other low--mass, spiral--dominated groups of galaxies in
absorption, with spectra of similar quality.
\end{abstract}

\section{Introduction}

That local intergalactic space could harbor substantial amounts of
hot, highly ionized gas, either as an extended halo surrounding the
Galaxy, or as an extended medium pervading the Local Group, was first
suggested by Spitzer (1956) \nocite{spitzer56} and by Kahn and Woltjer
(1959)\nocite{kahn59}. Such a medium would have a characteristic
temperature of order the virial temperature, estimated to be on the
order of $T \sim 2-3\times 10^6\rm\,K$, and be of very low density,
$n_e \sim 10^{-4}$~cm$^{-3}$ (\cite{maloney99}). Its emission, mainly 
soft thermal line emission from highly ionized metals and weak thermal
bremsstrahlung, would be very faint and extremely difficult to detect, 
and searches for diffuse hot gas have relied on detecting its
absorption lines in bright background continuum sources. Detailed
studies of Li-like C and O absorption lines in the UV in background
quasar spectra have indeed revealed the presence of gas in the
Galactic Halo ({\it e.g.}, \cite{savage00}) and, more recently, beyond
({\it e.g.}, \cite{tripp00}).

Intergalactic X-ray absorption spectroscopy has long been recognized
as having the potential of revealing more highly ionized gas, but the
required spectroscopic sensitivity has only recently become available
with the diffraction grating spectrometers on {\it Chandra} and
{\it XMM-Newton}. The first detection of resonance absorption lines
from H- and He-like oxygen and neon, in a spectrum of the 
bright BL Lac object PKS~2155$-$304 with the {\it Chandra} Low Energy
Transmission Grating Spectrometer, was recently reported
(\cite{nicastro02}).  
Here we report on multiple detections of the $1s-2p$ resonance lines
in H- and He-like oxygen in deep spectra of PKS~2155$-$304, Mkn 421,
and 3C~273, obtained with the Reflection Grating Spectrometer (RGS) on
{\it XMM-Newton}. The \ion{O}{7} lines appear at zero redshift, with
characteristic equivalent widths of 15~m\AA; the \ion{O}{7} line in
Mrk~421 appears marginally resolved, with a characteristic velocity
width of  $300\rm\,km\,s^{-1}$. We will argue that the bulk of the
absorption in these lines arises in an extended low density medium in
near collisional equilibrium, well outside our Galaxy, which is likely
the long sought-for intragroup medium of the Local Group.

\section{Data and Analysis}

Weak X-ray absorption features are detectable only when spectra
of sufficient photon statistics were acquired with the grating
spectrometers, and when various systematic effects of the detectors
to the histogrammed data sets can be minimized. To prepare such
spectra, we combined the available spectroscopic data toward the three
high latitude AGNs from a combination of calibration, performance
verification, and guaranteed time observation data sets available to
the {\it XMM-Newton} RGS team. Using all of the available datasets,
the effective exposure times added up to 412, 274 and 265 ksec for
3C~273, Mrk~421 and PKS~2155$-$304, respectively. Custom software was used
to operate on the observation data file (ODF) in each case, to produce
spectra aligned to the same dataspace grid. The custom software is
identical in function to, and shares a common origin with, the RGS
branch of the science analysis subsystem (SAS), except that it allowed
a flexible means by which to reduce systematic effects to the
data. Because there is a residual pointing uncertainty on the order of
2\arcsec\ for the analysis at this time, the reported velocities
should be accompanied by systematic wavelength uncertainties of order
of $\pm$5\rm m\AA, or $70\rm\, km\,s^{-1}$ for a line close to
20\AA. The relative error in wavelength between two line measurements
from the same observation is much lower.

\begin{figure}[p]
\centerline{
\psfig{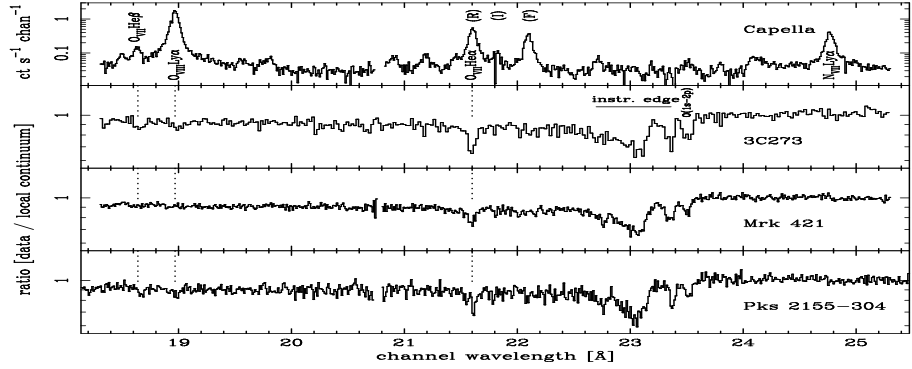}
}
\caption{A comparison of the three AGN spectra used in this study,
  displayed in the range 18.3 to 25.3\AA.\label{fig:abs_comparison}}  
\end{figure}

Because AGNs typically have variable spectra, the continuum spectrum 
from each observation was fit separately, and the ratio spectra
(``data:folded model'') were counts--weighted to yield a final
absorption spectrum toward each of the AGNs. These are displayed in
Figure~\ref{fig:abs_comparison}, uncorrected for instrumental and
interstellar absorption features. For comparison, the spectrum of
Capella is displayed on the top panel, which can be used to identify
the locations of strong transitions in the rest frame. Each AGN spectrum
exhibits a clear signature of intervening \ion{O}{7} by the $1s-2p$
He$\alpha$ transition at 21.60\AA. Hints of the \ion{O}{8} $1s-2p$
Ly$\alpha$ and \ion{O}{7} $1s-3p$ He$\beta$ transitions are also
seen at 18.63 and 18.97\AA, respectively. We fit the absorption line
profiles in the data within XSPEC using a Voigt model folded through
the instrumental redistribution matrices. 

\section{Results}

A quantitative summary of the absorption features measured is given
in Table~\ref{tab:results}. The spectrum toward Mrk~421 had superior
statistics, and provided detections in four transitions: $1s-2p$ in
\ion{O}{8}, \ion{O}{7}, \ion{Ne}{9} and \ion{C}{6}. Of these,
the strongest absorption feature (\ion{O}{7}) appears to be
marginally resolved, and yields an ionic velocity dispersion $\rm
V_{turb}$ between 200 and 560~$\rm km\,s^{-1}$, based on 90\%
confidence levels and, in a Gaussian approximation to the
line--of--sight velocity distribution characterized by $\rm \sigma_v$,
$\rm V_{turb} =  \sqrt{2}\,\sigma_v$. In most cases, fitting the
absorption features works significantly better when a line broadening
mechanism beyond the thermal velocity distribution is included.  

While the measured line equivalent widths (from Voigt profile fitting)
are given in Table~\ref{tab:results}, estimating the corresponding
ionic column densities ($\rm N_i$) requires an assumption of the
velocity distribution. The rightmost column gives a conservative lower
limit to the true column density for each ion species measured: We use
the 90\% lower limit to the equivalent width of the feature, and then
tabulate the corresponding ionic column density in the unsaturated
limit. The true column density, however, can depend strongly on the
true velocity distribution for a given measured equivalent width. For
example, for both Mrk~421 and 3C~273, the inferred $\rm N_i$ values
increase by a factor of 2 when the velocity dispersion parameter is
reduced to $100\rm\, km\,s^{-1}$ from $\sim300\rm\, km\,s^{-1}$, where
the fits provide marginally better $\chi^2$ values.

\begin{deluxetable}{lccccccc}
\tabletypesize{\scriptsize}
  \tablecaption{\label{tab:results} Absorption line fitting results.
  }
  \tablewidth{0pt}
  \tablehead{
    \colhead{Target} 
    & \colhead{EW\tablenotemark{a} (m\AA)}
    & \colhead{Significance}
    & \multicolhead{2}{Centroid Velocity\tablenotemark{a}}
    & \multicolhead{2}{$\rm V_{turb}$\tablenotemark{a}} 
    & \colhead{$\rm N_{i}(min)$\tablenotemark{b}} \\
    & 
    & 
    &    \colhead{min}&\colhead{max}
    &    \colhead{min}&\colhead{max}
    &    \colhead{$\rm 10^{16}\,cm^{-2}$}
  }

  \startdata
\\
\multicolumn{8}{c}{O~VII He$\alpha$ $1s-2p$ ($\lambda$21.602)}\\
  3C~273          & $26.3^{+4.5}_{-4.5}$ & $10.2\sigma$ & $-190$ &
  $+46$ & $0$ & $638$ & 0.76 \\
  Mrk~421         & $15.4^{+1.7}_{-1.7}$ & $15.7\sigma$ & $-254$&$-98$
  & $203$&$563$ & 0.48 \\
  PKS~2155        & $16.3^{+3.3}_{-3.3}$ & $8.1\sigma$  & $-182$&$+233$&
  $0$&$963$ & 0.45 \\
\\
\tableline
\\
\multicolumn{8}{c}{O~VIII Ly$\alpha$ $1s-2p$ ($\lambda$18.970)}\\
  3C~273          & $11.7^{+3.6}_{-3.7}$ & $5.6\sigma$ & $-47$&$+711$
  & $122$&$1020$ & 0.60 \\
  Mrk~421         & $4.3^{+1.0}_{-1.1}$  & $6.8\sigma$ & $-648$&$-366$
  & $0$&$387$ & 0.24 \\
  PKS~2155        & $9.0^{+2.6}_{-2.7}$  & $5.7\sigma$ & $-427$&$+95$
  & $415$&$1555$  & 0.48 \\
\\
\tableline
\\
\multicolumn{8}{c}{C~VI Ly$\alpha$ $1s-2p$ ($\lambda$33.734)}\\
  3C~273          & $12.9^{+4.1}_{-4.3}$ & $5.0\sigma$ & $-133$&$+111$
  & $0$ & $311$ & 0.21 \\
  Mrk~421         & $3.5^{+1.9}_{-1.6}$  & $3.6\sigma$ & $-164$&$+132$
  & $0$ & $292$ & 0.05 \\
  PKS~2155        & NA & NA & NA & NA
  & NA & NA & NA \\
\\
\tableline
\\
\multicolumn{8}{c}{Ne~IX He$\alpha$ $1s-2p$ ($\lambda$13.447)}\\
  3C~273          & $14.3^{+17.7}_{-14.3}$ & $2.3\sigma$ & NA & NA
  & NA & NA & NA \\
  Mrk~421         & $3.0^{+1.2}_{-1.3}$  & $3.7\sigma$ & -747 & 1138
  & NA & NA & 0.15 \\
  PKS~2155        & $0.0^{+2.8}_{-0.0}$ & $0.0\sigma$ & NA & NA
  & NA & NA & NA \\
  \tablenotetext{a}{Ranges and errors given are 90\% confidence
  limits. Velocities are given in $\rm km\,s^{-1}$. Absolute velocity
  uncertainties due to (systematic) pointing uncertainties are not
  reflected in these confidence intervals, and are expected to be on
  the order of $\Delta v \sim 70\times(\lambda/20)^{-1}\rm\,km\,
  s^{-1}$.} 
  \tablenotetext{b}{Minimum ion column density is computed by
  converting the EW lower limit in the unsaturated approximation:
  $\rm N_i(min) \equiv EW_{min}\times  
  (f\, \pi e^2 /  mc^2)^{-1} \lambda^{-2}$.}
  \enddata
\end{deluxetable}


Another fact to note in the fitting results is that the
\ion{O}{7} and \ion{O}{8} feature positions do not match up in
their systemic velocities and velocity widths. While there are in most
cases systemic velocities and $\rm V_{turb}$ that satisfy the 90\% limits
for both lines for a given AGN, the differences seen in them evoke a
possible situation where the absorption features may probe different
velocity components, and therefore different phases of the
intervening gas. When a fit was performed simutaneously to the
\ion{O}{7} and \ion{O}{8} features with velocity parameters tied,
a lower equivalent width and column density was derived for the
\ion{O}{8} feature, because the fit was driven by details of the
\ion{O}{7} feature profile -- additional velocity components are
then required to fit the excess \ion{O}{8}. 
From the column density ratios for \ion{O}{8} to \ion{O}{7}
derived this way (0.3 for Mrk~421 to 0.8 for PKS~2155) we estimate the
electron temperatures by assuming collisional equilibrium and that the
\ion{O}{7} and \ion{O}{8} occupy the same volume. The electron
temperature range, according to this assumption, is $2-5\times
10^{6}$~K, which brackets the 1$\sigma$ confidences for each of the
targets. 

\begin{figure}[h]
  \centerline{
    \psfig{figure=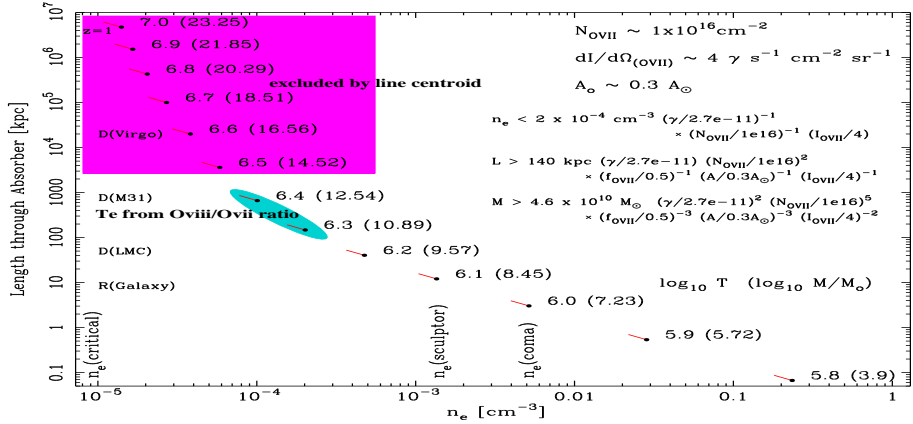,width=\textwidth,height=0.29\textheight,angle=-90}
  }
  \caption{Solutions to the emitter--absorber problem for a uniform
    medium under CIE conditions.  For each temperature, the
    black dot represents the solution for $n_{\rm e}$ and $L$, while the 
    red line is along the column density constraint, pointing away from
    the upper limit for $n_{\rm e}$.
    \label {fig:l_ne_m_constraints}} 
\end{figure}

Collisional excitation and emission from the same medium
may be estimated from the ROSAT all--sky survey toward
these lines of sight. Alternatively, a better estimate of the emission
line contribution to the diffuse X--ray background is provided by
results of a recent sounding rocket experiment (X--ray Quantum
Calorimeter, \cite{mccammon02}). Their analysis estimated that the 
Oxygen line fluxes (for $z<0.01$) together account for 32\% of the R4
band count rates. However, because the calorimeter provided individual
line fluxes ({\it e.g.}, $\rm dI/d\Omega = 4.8 \pm 0.8\rm\,
phot\,s^{-1}\,cm^{-2}\,sr^{-1}$ in the \ion{O}{7} triplet) electron
densities and lengths through the emitter--absorber may be estimated.
In the approximation that the medium has a uniform density and
temperature out to some distance $L$, the ionic column density should
be $N_{\rm i} = A_{\rm O}\,f_{\rm i}\,n_{\rm p}\,L$ and the diffuse
emission intensity ${\rm d}I/{\rm d}\Omega = (1/4\pi) A_{\rm O}\,f_{\rm i}\,
n_{\rm p}\,n_{\rm e}\,\gamma(T) L$ where $A_{\rm O}$, $f_{\rm i}$ and
$\gamma(T)$ are the elemental abundance, ionic fraction and
collisional rate coefficient for the transition, respectively. Since
the ion column density is formally a lower limit and the surface
brightness is an upper limit to true diffuse contribution, an upper
limit to the electron density $n_{\rm e}$ can be estimated in a way
that is independent of oxygen abundance. Then, by assuming an oxygen
abundance and ionic fraction, a lower limit to the length $L$ is
estimated. Figure~\ref{fig:l_ne_m_constraints} shows a graphical
solution to the emitter--absorber for a range of temperatures and
some assumptions ($A_{\rm O} \sim 0.3 A_{\odot}$,
$f_{\rm i} \sim 0.5$, ${\rm d}I/{\rm d}\Omega \sim 4\,\rm
phot\,s^{-1}\,cm^{-2}\,sr^{-1}$ and $N_{\rm i} \sim 10^{16}\,\rm
cm^{-2}$). From the temperature $T$ inferred from the column
density ratios, the electron density estimate yields $n_{\rm e} <
2\times 10^{-4}\,\rm cm^{-3}$ and a length scale $L > 140\,{\rm
  kpc}\,(A_{\rm O}/0.3\,A_\odot)^{-1}$, $\sim$~40 times larger (and
more tenuous) than the \ion{O}{6} absorbing Galactic halo
(\cite{savage00}).  


\section{Conclusions}

The results of this initial survey of $z \sim 0$ absorption by highly
ionized gas in the high Galactic latitude pointings provide a picture
of IGM in the local group that can be tested toward other galaxies in
poor groups. With impact parameters on the order of
100~kpc from spiral galaxies such as our own, the spectra of more
distant AGNs should contain absorption features with characteristic
column densities of $N_{\rm i} \sim 10^{16}\,\rm cm^{-2}$ in 
\ion{O}{7}. Such extended, tenuous halos or intragroup gas
should consequently be accompanied by a diffuse glow, that would only
double the local surface brightness of the X--ray background's diffuse
and line--rich component on the 100~kpc scale, and would be very
difficult to detect.

\bibliographystyle{apalike}
\chapbblname{rasmussen_igmc}
\chapbibliography{rasmussen_igmc}


\end{document}